\documentstyle[preprint,pra,aps]{revtex}
\newcommand{\be}{\begin{equation}}
\newcommand{\ee}{\end{equation}}
\newcommand{\ba}{\begin{eqnarray}}
\newcommand{\ea}{\end{eqnarray}}

\draft

\begin{document}
\draft
\title{Thermodynamics' 0-th-Law in a nonextensive scenario}
\author{S. Mart\'{\i }nez$^{1,\,2}$ \thanks{%
E-mail: ~martinez@venus.fisica.unlp.edu.ar}, F. ~Pennini$^{1,2}$ \thanks{%
E-mail:~pennini@venus.fisica.unlp.edu.ar}, and A. ~Plastino$^{1,\,2}$
\thanks{%
Corresponding Author, E-mail:~plastino@venus.fisica.unlp.edu.ar }}
\address{$^1$Departamento de F\'{\i}sica, Facultad de Ciencias Exactas,\\
Universidad Nacional de La Plata (UNLP),\\
C.C. 727, 1900 La Plata, Argentina.\\
$^2$ Instituto de F\'\i sica\\
La Plata (IFLP), CONICET-UNLP.}
\maketitle

\begin{abstract}
Tsallis' thermostatistics \cite
{mendes,review,rajagopal,boghosian,abe,grigolini,naudts,landsberg,chimento,t01,t1,t03,t04,t05}
is by now recognized as a new paradigm for statistical mechanical
considerations. However, the generalization of thermodynamics'
zero-th law to a nonextensive scenario is plagued by difficulties
\cite{review}. In this work we suggest  a way to overcome this
problem.

\vspace{0.2 cm}  PACS: 05.30.-d,05.70.Ce.

\vspace{1 cm}
\end{abstract}

\newpage

\section{Introduction}

Tsallis' thermostatistics offers a suitable and quite significant
generalization of the Boltzmann-Gibbs statistical mechanics, that has found
multiple applications \cite
{mendes,review,rajagopal,boghosian,abe,grigolini,naudts,landsberg,chimento,t01,t1,t03,t04,t05}%
. However, it can not yet comfortably deal with thermodynamics'
zero-th law, as pointed out by Tsallis himself in \cite{review}.
In \cite{abe2} Abe has advanced an interesting, tentative solution
to the zero-th law conundrum with reference to the micro-canonical
analysis of a system composed of two-subsystems in thermal
equilibrium. Such an analysis leads to the appearance of
temperatures that {\it depend upon the nonextensive partition
function} $\bar{Z}_q$,where $q$ is the Tsallis' non-extensivity
index.

As shown by the present authors in a recent study \cite{newnorm},
one can recast Tsallis' variational problem (using normalized
expectation values) in such a manner that the extremum one thereby
finds is guaranteed to correspond to a maximum (and not to other
types of extrema) of Tsallis information measure
\begin{equation}
\frac{S_q}{k} = \frac{1-Tr(\hat{\rho}^q)}{q-1},  \label{entropia}
\end{equation}
 ($\hat \rho$ is
the density operator and $k$ is the Boltzmann constant, or more
generally, the information unit)  because the associated Hessian
is diagonal. This treatment involves a new set of Lagrange
multipiers $\lambda_j^{\prime}$, to be referred to as the
``Optimal set" (OLM), different, but related, to the original
Tsallis-Mendes-Plastino (TMP) one ($\lambda_j$'s) \cite{mendes}

\begin{equation}  \label{lambda'}
\lambda_j^{\prime }= \frac{\lambda_j}{\bar{Z}_{q}^{1-q}},
\end{equation}
where the partition function $\bar{Z}_q$ is involved.

Another interesting work in this context is that of Ref. \cite{rama}, where
it is shown that for those particular systems whose partition function is
given by $Z_{BG}\propto l^a (\beta^{\prime})^{-a}$ ($a$ is a dimensionless
parameter, and $l$ is a characteristic length), the inverse
(thermodynamical) temperature  becomes associated with {\it our} $%
\beta^{\prime}$ and not with the TMP $\beta$.

In the present effort we tackle the vexing zero-th law problem starting with
the working hypothesis that $1/\beta^{\prime}$ {\it is} indeed the
temperature. We show that such a hypothesis  reconciles Tsallis' formalism
with the zero-th law. A price has to be paid, of course.  The important
relations \cite{mendes}
\begin{eqnarray}
\frac{\partial}{\partial E} \left( \frac{S_q}{k} \right) &=& \beta
\label{termo11} \\
\frac \partial {\partial \beta}\left( \ln _q Z_q\right) &=&- E,
\label{termo22}
\end{eqnarray}
{\it lose} their basic character, because $\beta$ depends upon the
partition function. They are however recovered in the $q
\rightarrow 1$ limit.

We shall tackle the zero-th law problem starting
with the hypothesis of \cite {abe2}: one deals
with the Hamiltonian of a system composed of two
independent subsystems (in the sense that their
mutual interaction is negligible). The system's
density operator is product of those pertaining to
the subsystems. Before, however, a short
recapitulation is necessary.

\section{Main results of the OLM formalism}

The most general quantal treatment is made in a basis-independent way, which
requires consideration of the statistical operator (or density operator) $%
\hat{\rho}$ that maximizes Tsallis' entropy, subject to the foreknowledge of
$M$ generalized expectation values (corresponding to $M$ operators $\widehat{%
O}_j$).

Tsallis' normalized probability distribution \cite{mendes} is
obtained by following the well known MaxEnt route \cite{katz}.
Instead of effecting the
variational treatment of \cite{mendes}, involving Lagrange multipliers $%
\lambda_j$, we pursue the alternative path developed in \cite{newnorm}, with
Lagrange multipliers $\lambda^{\prime}_j$. One maximizes Tsallis'
generalized entropy (\ref{entropia}) \cite{t01,t1,t2} subject to the
constraints (generalized expectation values) \cite{t01,newnorm}

\begin{eqnarray}
Tr(\hat{\rho}) &=&1 \\
Tr\left[ \hat{\rho}^q\left( \widehat{O}_j- \left\langle\widehat{O}%
_j\right\rangle _q\right) \right] &=&0,  \label{vinculos}
\end{eqnarray}
where $\widehat{O}_j$ ($j=1,\ldots ,M$) denote the M relevant observables
(the observation level \cite{aleman}), whose generalized expectation values
\cite{mendes}

\begin{equation}  \label{gener}
\left\langle\widehat{O}_j \right\rangle _q = \frac{Tr(\hat \rho^q \widehat{O}%
_j)}{Tr(\hat \rho^q)},
\end{equation}
are (assumedly) a priori known. The resulting density operator reads \cite
{newnorm}

\begin{equation}
\hat{\rho}=\bar{Z}_q^{-1}\left[ 1-(1-q)\sum_j^M\,\lambda _j^{\prime }\left(
\widehat{O}_j- \left\langle \widehat{O}_j\right\rangle _q\right) \right] ^{%
\frac 1{1-q}},  \label{rho}
\end{equation}
where $\bar{Z}_q$ stands for the partition function

\begin{equation}
\bar{Z}_q=Tr\left[ 1-(1-q)\sum_j\lambda _j^{\prime }\left( \widehat{O}%
_j-\left\langle \widehat{O}_j\right\rangle_q\right) \right] ^{\frac 1{1-q}}.
\label{Zqp}
\end{equation}

It is shown in \cite{newnorm} that
\begin{equation}
Tr(\hat{\rho}^q)=\bar{Z}_q^{1-q},  \label{relac1}
\end{equation}
and that Tsallis' entropy can be cast as

\begin{equation}
S_q=k\;{\rm \ln }_q\bar{Z}_q,  \label{S2}
\end{equation}
with ${\rm \ln }_q\bar{Z}_q=(1-\bar{Z}_q^{1-q})/(q-1)$. These results
coincide with those of TMP \cite{mendes} in their normalized treatment. If,
following \cite{mendes}, we define now
\begin{equation}
\ln _q Z_q={\rm \ln }_q\bar{Z}_q- \bar{Z}_{q}^{1-q}\sum_j\lambda
_j^{\prime }\ \left\langle \widehat{O}_j \right\rangle _q,
\label{lnqz'}
\end{equation}
we are straightforwardly led to \cite{newnorm}
\begin{eqnarray}
\frac{\partial}{\partial \left\langle \widehat{O}_j \right\rangle
_q}\left( \frac{S_q}{k} \right) &=&\bar{Z}_{q}^{1-q}\lambda
_j^{\prime } \label{termo1} \\ \frac \partial {\partial \lambda
_j^{\prime }}\left( \ln _qZ_q\right) &=&-\bar{Z}_{q}^{1-q}
\left\langle \widehat{O}_j \right\rangle _q. \label{termo2}
\end{eqnarray}

Equations (\ref{termo1}) and (\ref{termo2}) are modified Information Theory
relations of the type that one uses to build up, {\em \`a la} Jaynes \cite
{katz}, Statistical Mechanics. The basic Legendre-structure relations (of
which (\ref{termo11}) and (\ref{termo22}) are the canonical example) can be
recovered in the limit $q \rightarrow 1$.

As a special instance of Eqs. (\ref{termo1}) and (\ref{termo2}) let us
discuss the Canonical Ensemble, where they adopt the appearance
\begin{eqnarray}
\frac{\partial }{\partial U_q}\left(\frac{S_q}{k}\right)&=&\bar{Z}%
_{q}^{1-q}\beta ^{\prime }  \label{C1} \\ \frac \partial {\partial
\beta ^{\prime }}\left( \ln _qZ_q\right) &=&-\bar{Z}_{q}^{1-q}U_q,
\label{C2}
\end{eqnarray}
with (see equation (\ref{lnqz'}))
\begin{equation}
\ln _qZ_q=\ln _q\bar{Z}_q-\beta ^{\prime }U_q.  \label{lnqz'can}
\end{equation}

Equations (\ref{C1}) and (\ref{C2}) can be translated into (\ref{termo11})
and (\ref{termo22}) via (\ref{lambda'}).

\section{Thermodynamical equilibrium}

We tackle now the question we wish to address in this effort: to discuss
anew the problem of thermodynamical equilibrium on the basis of the results
of the preceding Section. Let us consider a composed isolated Hamiltonian
system $A+B$, within the framework of the Microcanonical Ensemble. These two
subsystems interact via heat exchange.

Following Gibbs, we make the usual assumptions \cite{review}:

\begin{enumerate}
\item  $\label{a1}$The interaction energy is negligible
\begin{equation} \label{hypo1}
\widehat{{\cal H}}(A+B)\sim \widehat{{\cal H}}(A)+\widehat{{\cal H}}(B).
\end{equation}

\item  \label{a2}The subsystems $A$ and $B$ are essentially independent in
the sense of the theory of probabilities, i.e.
\begin{equation}
\hat{\rho}(A+B)\sim \hat{\rho}(A)\hat{\rho}(B).  \label{rhosuma}
\end{equation}
\end{enumerate}

The energy distributions are here given, for each system, by specializing (%
\ref{rho}) and (\ref{Zqp}) to the instance $M=1$ and $\hat O^{(G)}_1 \equiv
\widehat{{\cal H}} (G),\,\,\,G=A,\,B$. It easily follows from Eq. (\ref
{gener}) that \cite{review}

\begin{equation}
U_q(A+B)=U_q(A)+U_q(B).  \label{uu}
\end{equation}

Now, after a bit of algebra Eq. (\ref{entropia}) yields (pseudo-additivity
\cite{review})

\begin{equation}
\frac{S_q(A+B)}{k(A+B)}=\frac{S_q(A)}{k(A)}+\frac{S_q(B)}{k(B)}+(1-q)\frac{%
S_q(A)}{k(A)}\frac{S_q(B)}{k(B)},  \label{su}
\end{equation}
which we here recast in the fashion

\begin{eqnarray}
&&\frac{{\rm ln}\left[ 1+\frac{(1-q)S_{q}(A+B)}{k(A+B)}\right] }{1-q}=
\label{ss} \\
&&=\frac{{\rm ln}\left[ 1+\frac{(1-q)S_{q}(A)}{k(A)}\right] }{1-q}+\frac{%
{\rm ln}\left[ 1+\frac{(1-q)S_{q}(B)}{k(B)}\right] }{1-q},  \nonumber
\end{eqnarray}
where we have made it explicit the fact that the constant $k$ could,
eventually, depend upon the system's nature.

Focus attention now upon Eqs. (\ref{uu}) and (\ref{ss}). For a closed
system, both energy and entropy are conserved. As a consequence:

\begin{eqnarray}
\delta U_q(A) & = & -\delta U_q(B),  \label{qu} \\
\frac{1}{ Tr[\rho(A)]^q}\delta\left(\frac{ S_q(A)}{k(A)}\right)&=& - \frac{1%
}{ Tr[\rho(B)]^q}\delta\left(\frac{ S_q(B)}{k(B)}\right) .  \label{zx}
\end{eqnarray}

Introduction of (\ref{relac1}) into (\ref{zx}) yields now

\begin{equation}
\frac{1}{\bar{Z}_q(A)^{1-q}}\delta\left(\frac{S_q(A)}{k(A)}\right)= - \frac{1%
}{\bar{Z}_q(B)^{1-q}}\delta\left(\frac{S_q(B)}{k(B)}\right).  \label{qa}
\end{equation}

The next step is to consider the ratio between (\ref{qa}) and (\ref{qu}),
keeping in mind (\ref{C1}). One immediately finds the equality

\begin{equation}
\beta^{\prime}(A)=\beta^{\prime}(B),
\end{equation}
i.e., if we set $\beta^{\prime}\propto 1/T$, thermal equilibrium
between $A$ and $B$ arises in a natural fashion and the {\it
thermodynamics' zero-th law} is obtained. This constitutes the
essential result of the present effort. Notice that one assumes
here that $\beta^{\prime}$, not $\beta$ (as in \cite{abe2}), is
proportional to $\frac{1}{T}$, a fact first observed in
\cite{rama} for those special systems whose partition function is
of the form $Z_{BG}\propto l^a (\beta^{\prime})^{-a}$, with $a$ a
dimensionless parameter, and $l$ a characteristic length.

In terms of $\beta$ (the TMP Lagrange multiplier) we have

\begin{equation}
\frac{\beta(A)}{\bar{Z}_q^{1-q}(A)}=\frac{\beta(B)}{\bar{Z}_q^{1-q}(B)}.
\end{equation}
To take $\beta$ as proportional to $\frac{1}{T}$ forces one to work with a
temperature that depends upon the partition function \cite{abe2}.

The present work shows that one can reconcile the zero-th law with
Tsallis' thermostatistics without going to the limit $q
\rightarrow 1$. In order to assess to what an extent have we
succeeded it remains to ascertain the self-consistency of the
Gibbs' hypothesis  (\ref{hypo1},\ref{rhosuma}) within our
nonextensive framework. We reconsider the application of Eq.
(\ref{rho}) to our present situation and define

\begin{equation}  \label{fa}
\widehat F(A) = \left[1-(1-q) \beta \left( \widehat{{\cal H}}%
(A)-U_{q}(A)\right) \right] ^{\frac{1}{1-q}},
\end{equation}
with a similar expression for $\widehat F(B)$. We have then
\begin{equation}
\hat{\rho}(A)\hat{\rho}(B) = \frac{\widehat F(A)}{\bar{Z}_{q}(A)}\,\, \frac{%
\widehat F(B)}{\bar{Z}_{q}(B)},
\end{equation}
which, after explicit evaluation, and keeping just first order terms gives

\begin{eqnarray}
&&\hat{\rho}(A+B)=\hat{\rho}(A)\hat{\rho}(B)-  \label{rhoAB} \\
&&-(1-q)\beta ^{2}\frac{[\widehat{{\cal H}}(A)-U_{q}(A)]}{\bar{Z}_{q}(A)}%
\frac{[\widehat{{\cal H}}(B)-U_{q}(B)]}{\bar{Z}_{q}(B)},  \nonumber
\end{eqnarray}

\noindent which is the promised result. As pointed out in
\cite{raggio}, our subsystems are not exactly independent. But the
last term on the r.h.s. of the above expression is negligible for
i) high temperatures, ii) the thermodynamic limit (see below), or,
of course, for iii) $q$-values close to unity.

\section{Conclusions}

We have carefully reconsidered the validity of the zero-th law of
thermodynamics in a Tsallis' environment. It has been shown to remain
approximately valid.

The question revolves around the independence of two independent subsystems
and $A$, $B$ that are brought into thermal contact. We have found that they
can indeed be regarded as independent in quite important instances:

\begin{itemize}
\item  1) $q\rightarrow 1$, of course,

\item  2) in the high temperature limit, and

\item  3) {\it for systems in contact with a heat reservoir}, because, if $A$%
, say, is the reservoir, $[\widehat{{\cal H}}(A)-U_{q}(A)]$ is a null
operator (the mean energy of a reservoir coincides, by definition, with one
of its eigenenergies \cite{deslogue}). Now you invoke implicitly the heat
reservoir notion whenever you use a thermometer!
\end{itemize}

Summing up, for practical purposes the zero-th law of thermodynamics is
valid in a Tsallis scenario.

\acknowledgements The financial support of the National Research Council
(CONICET) of Argentina is gratefully acknowledged. F. Pennini acknowledges
financial support from UNLP, Argentina.

\end{document}